\documentclass[reprint,showpacs,preprintnumbers,nofootinbib,amsmath,amssymb,aps,prd,floatfix]{revtex4-1}
\pdfoutput=1
\usepackage{graphicx}
\usepackage{bm}
\usepackage{subfigure}
\usepackage{url}
\usepackage[hyperindex]{hyperref}
\usepackage{color}
\usepackage[ddmmyy,24hr]{datetime}
\usepackage{bigdelim}
\usepackage{booktabs}
\usepackage{dcolumn}
\usepackage{multirow}
\usepackage{subfigure}
\usepackage{cancel}
\usepackage{stackrel}
\usepackage{paralist}
\usepackage{xspace}
\usepackage[utf8]{inputenc}
\newcommand{\nua}[1]{\ensuremath{\rlap{\kern-2.5pt\ensuremath{\overset{\scriptscriptstyle(-)}{\phantom{\nu}}}}{\ensuremath{{\nu}_{#1}}}}}

\renewcommand{\P}{\ensuremath{{P}}}

\newcommand{\QW}{$Q_{W}$}

\usepackage{enumitem}

\makeatletter
\def\namedlabel#1#2{\begingroup
    #2%
    \def\@currentlabel{#2}%
    \phantomsection\label{#1}\endgroup
}

\makeatother
\begin{document}

\title{Reinterpreting the weak mixing angle from atomic parity violation in view of the Cs neutron rms radius measurement from COHERENT}

\author{M. Cadeddu}
\email{matteo.cadeddu@ca.infn.it}
\affiliation{INFN, Sezione di Cagliari,
Complesso Universitario di Monserrato - S.P. per Sestu Km 0.700,
09042 Monserrato (Cagliari), Italy}

\author{F. Dordei}
\email{francesca.dordei@cern.ch}
\affiliation{INFN, Sezione di Cagliari,
Complesso Universitario di Monserrato - S.P. per Sestu Km 0.700,
09042 Monserrato (Cagliari), Italy}

\begin{abstract}

Using the model independent average neutron rms radius of $^{133}\text{Cs}$ and $^{127}\text{I}$ obtained from the analysis of the coherent elastic neutrino-nucleus scattering data of the COHERENT experiment, we remove the long-standing $1.5\,\sigma$ tension between the Standard Model prediction and the weak mixing angle measurement from the atomic parity violation (APV) in caesium. The updated APV result becomes $\sin^2 \vartheta_{\text{W}}=0.239{}^{+0.006}_{-0.007}$, to be compared with the  Standard Model prediction at low momentum transfer, $\sin^2 \vartheta_{\text{W}}^{\textrm{SM}} = 0.23857(5)$. Moreover, exploiting the fact that the APV result is highly sensitive to the caesium neutron rms radius, $R_{n}$, and assuming that the Standard Model is correct, we combine the APV and the COHERENT measurements in order to get a better determination of $R_{n}$. The value of \mbox{$R_{n}=5.42\pm 0.31\,\text{fm}$} is obtained, improving significantly the current uncertainty. This result allows to infer a meaningful value of the caesium neutron skin, the difference between the neutron and proton distribution radii, equal to $\Delta R_{np}=0.62\pm 0.31\,\text{fm}$, showing for the first time a $2\, \sigma$ deviation from zero.

\end{abstract}


\maketitle

The weak mixing angle, $\vartheta_{\text{W}}$, also known as the Weinberg angle, is a fundamental parameter in the theory of the electroweak (EW) interactions included in the Standard Model (SM) of particle physics~\cite{PDG2018, Kumar:2013yoa}. In practice, the quantity $\sin^2 \vartheta_{\text{W}}$ is usually quoted instead of the weak mixing angle itself.
 
Since the 1970s, the predictions of the SM EW theory have been extensively tested thanks to low-energy measurements in neutrino scattering and deep inelastic polarized electron-deuteron scattering~\cite{PDG2018,Wang:2014bba}. From then, it took almost one decade to have the first precision measurement of $\sin^2 \vartheta_{\text{W}}$ by means of dedicated neutrino and charged lepton scattering experiments. Subsequently, the experiments performed at the LEP1 and SLC colliders also provided valuable information on the state of health of the SM, confirming the EW theory predictions thanks to many measurements of Z-boson properties, finally establishing the SM as the correct theory. In particular, at the Z-pole, it was possible to achieve the most precise measurements of $\sin^2 \vartheta_{\text{W}}$ in the high-energy EW sector, in perfect agreement with SM predictions. 

\begin{figure}[!t]
\centering
\includegraphics*[width=\linewidth]{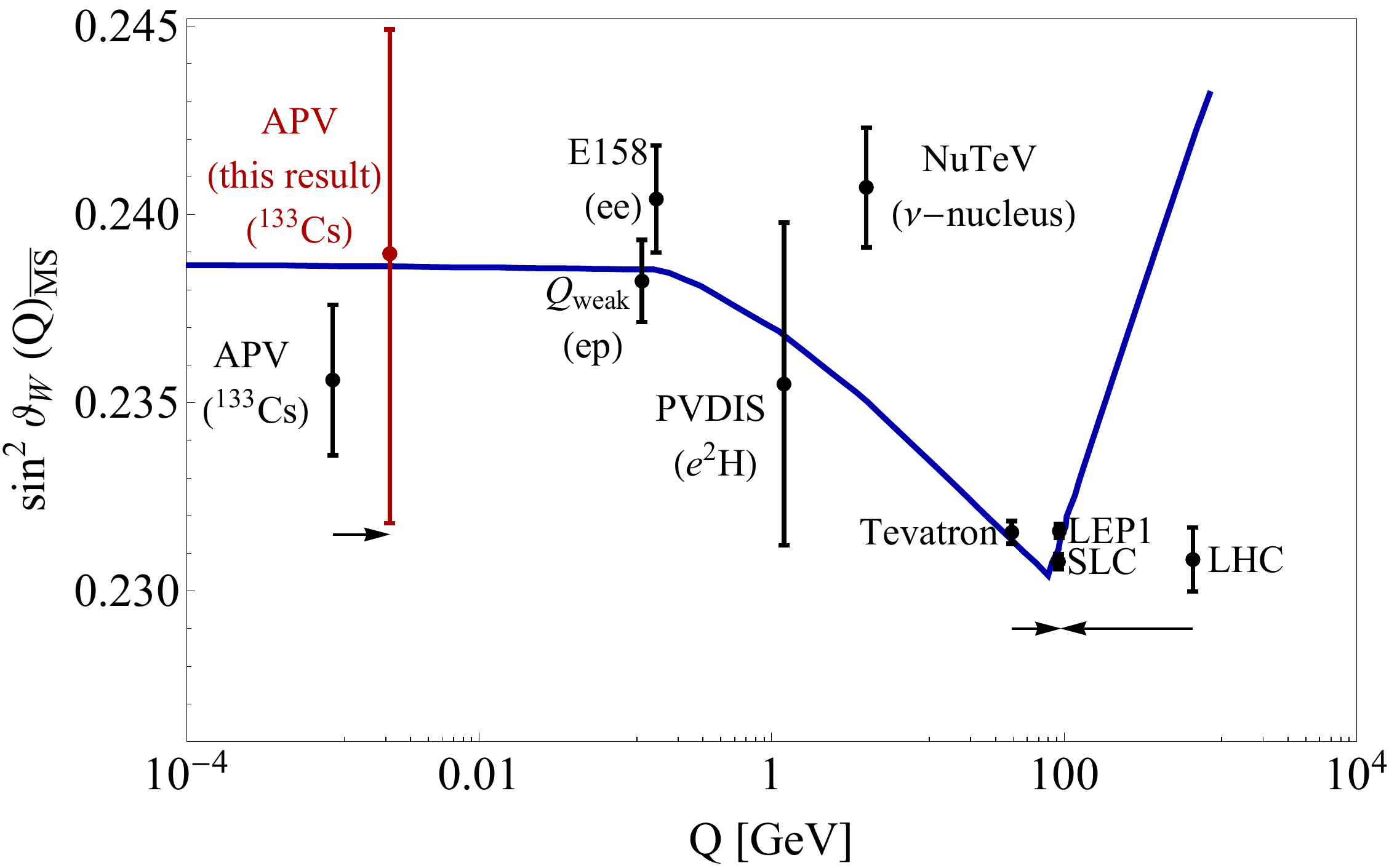}
\caption{ \label{fig:running}
Variation of $\sin^2 \vartheta_{\text{W}}$ with energy scale Q. The SM prediction is shown as the solid curve, together with
experimental determinations in black at the $Z$-pole~\cite{PDG2018} (Tevatron, LEP1, SLC, LHC),
from APV on caesium~\cite{Wood:1997zq,PhysRevLett.109.203003}, which has a typical momentum transfer given by $\langle Q\rangle\simeq$~2.4 MeV, M{\o}ller scattering~\cite{Anthony:2005pm} (E158), deep inelastic scattering of polarized electrons on deuterons~\cite{Wang:2014bba} ($ e^2H $ PVDIS) and from
neutrino-nucleus scattering~\cite{Zeller:2001hh} (NuTeV) and the new result from the proton's weak charge 
at $Q = 0.158$ GeV~\cite{Androic:2018kni} ($ Q_{weak} $). In red it is shown the result derived in this paper, obtained correcting the APV data point by the direct caesium neutron rms radius determination obtained in Ref.~\cite{PhysRevLett.120.072501}. For clarity we displayed the old APV point to the left and the Tevatron and LHC points horizontally to the left and to the right, respectively.}
\end{figure}

The experimental determination of $\sin^2 \vartheta_{\text{W}}$ provides a direct probe of physics phenomena not included in the SM, usually referred to as new physics. A summary of the weak mixing angle measurements as a function of the energy scale, Q, is shown in Fig.~\ref{fig:running}, along with the SM predicted running of $\sin^2 \vartheta_{\text{W}}$, calculated in the so-called modified minimal subtraction ($\overline{ MS }$) renormalization scheme~\cite{PDG2018, Erler:2004in,Erler:2017knj}.  
The most recent experimental value, which falls in the mid-energy range, has been derived from the measurement of the weak charge of the proton, $Q^{p}_{W}$, performed by the $Q_{\text{weak}}$ Collaboration and found to be $Q^{p}_{W}=0.0719\pm0.0045$~\cite{Androic:2018kni}, showing an excellent agreement with the SM prediction. 
Moving to the low-energy sector, the most precise weak angle measurement so far belongs to the so-called atomic parity violation (APV) experiments, also known as parity nonconservation (PNC), using caesium atoms. Atomic parity violation is caused by the weak interaction, either by $Z$-boson exchange between the electrons and the nucleus or by {\P}-violating inter-nuclear forces, and it is manifested in {\P}-violating atomic observables~\cite{Roberts:2014bka}. Such experiments plays a unique role complementary to those at high-energy~\cite{Safronova:2017xyt}.
In particular, APV is highly sensitive to extra $Z$ ($Z'$) bosons predicted in grand unified theories, technicolor models, supersymmetry and string theories, underscoring the need for improved experimental determinations of $\sin^2 \vartheta_{\text{W}}$ in the low-energy regime~\cite{Safronova:2017xyt}. 
Moreover, historically the APV measurement has moved significantly over the years, being mostly lower than the SM prediction, at near zero momentum transfer, calculated in the $\overline{ MS }$ scheme~\cite{PDG2018}
\begin{equation}
\label{sin2theta}
\sin^2\theta_W^{\rm SM} = 0.23857(5)\,,
\end{equation}
motivating a further investigation of all the inputs entering in this measurement.

The APV determination of $\sin^2 \vartheta_{\text{W}}$ is derived by measuring the weak charge of $^{133}\text{Cs}$, $Q^{\text{Cs}}_W$. In the SM, for a nucleus with $N$ neutrons and $Z$ protons, the weak charge including EW corrections is defined as~\cite{Erler:2013xha}
\begin{align}
\label{QWSMEWcorrections1}
Q_{{W}}^{\rm SM+rad. corr.} &\equiv
- 2 [ Z (g_{AV}^{\, e p} + 0.00005) \\\nonumber
 &+ N (g_{AV}^{\, e n} + 0.00006) ] 
\left( 1 - {\alpha\over 2 \pi} \right) \\\nonumber
&\approx Z (1 - 4 \sin^2 \theta_W^{\rm SM}) - N, 
\end{align}
where $\alpha$ is the fine structure constant and the nucleon couplings, $g_{AV}^{\, e p}$ and $g_{AV}^{\, e n}$, are given by
\begin{equation}
g_{AV}^{\, e p}  \approx - {1\over 2} + 2 \sin^2\theta_W^{\rm SM}, \,\,\,\,\mathrm{and} \,\,\,\, g_{AV}^{\, e n}  \approx {1\over 2}\,.
\end{equation}
The numerically small adjustments in Eq.~(\ref{QWSMEWcorrections1}) are discussed in Ref.~\cite{Erler:2013xha} and include the result of the $\gamma Z$-box correction from Ref.~\cite{PhysRevLett.109.262301}. 
For caesium, where $N=78$ and $Z=55$, the SM prediction of the weak charge is~\cite{PDG2018}
\begin{equation}
\label{eq:QWSM}
Q_W^{\rm SM+rad. corr.}= -73.23(1)\,.
\end{equation}

Experimentally, the weak charge of a nucleus is extracted from the ratio of the parity violating amplitude, $E_{PNC}$, to the Stark vector transition polarizability, $\beta$, and by calculating theoretically $E_{\rm PNC}$ in terms of \QW, leading to
\begin{equation}
\label{QWeq}
Q_W= N \left( {{\rm Im}\, E_{\rm PNC}\over\beta} \right)_{\rm exp.} 
\left( {Q_W \over N\, {\rm Im}\, E_{\rm PNC}} \right)_{\rm th.} \beta_{\rm exp.+th.}\,,
\end{equation}
where $\beta_{\rm exp.+th.}$ and $(\mathrm{Im}\, E_{\rm PNC})_{\rm th.}$ are determined from atomic theory, and Im stands for imaginary part.
In 1997, the most precise result of $Q^{\rm Cs}_W$ was obtained using the experimental input~\cite{Wood:1997zq}
$({\rm Im}\, E_{\rm PNC}/{\beta})_{\rm exp}= - 	1.5935(56)~{\rm mV/cm}$ or 
$({\rm Im}\, E_{\rm PNC}/{\beta})_{\rm exp} = - 3.0988(109) \times 10^{-13} |e|/a_B^2$ (where $a_B$ is the Bohr radius and $|e|$ is the electric charge), if $\beta$ is given in atomic units to be consistent with Eq.~(\ref{QWeq}).
In 1999, a more precise value of $Q^{\rm Cs}_W$ was extracted, using the most accurate value of $\beta$ at the time, $\beta = 26.957(51)\, a_B^3$ coming from an analysis~\cite{Dzuba:2000gf} of the Bennett and Wieman measurements~\cite{Bennett:1999pd}.
Moreover, the theoretical uncertainty of $({\rm Im}\, E_{\rm PNC})_{\rm th.}$ was re-evaluated by improving the calculation with the comparison with other measurable quantities, such as hyperfine levels, obtaining
\begin{equation}
\label{ImEPNC}
({\rm Im}\, E_{\rm PNC})_{\rm th.}=0.9065(36)\times10^{-11}|e|a_B^2 \frac{Q_W}{N} \,.
\end{equation}
Using this input, the value of \mbox{$Q^{\rm Cs}_W=-72.06~(28)_{\rm exp.}~(34)_{\rm th.}$} was measured, which differed from the SM prediction at the time by $2.3\, \sigma$.
Over the past decade, several theoretical developments appeared to reduce the tension with  the SM (such as the inclusion of Breit and QED radiative corrections), shifting the numerical coefficient in Eq.~(\ref{ImEPNC}) to $0.8906(26)\times10^{-11}$. This led to \mbox{$Q^{\rm Cs}_W = -73.16(29)_{\rm exp.}~(20)_{\rm th.}$}, in excellent agreement with the SM expectation.  
However, a recent re-evaluation~\cite{PhysRevLett.109.203003}, with the inclusion of many-body effects that were neglected in previous works, moved back the result to values more similar with earlier works~\cite{Dzuba:2002kx}, namely $({\rm Im}\, E_{\rm PNC})_{\rm th.}=0.8977(40)\times10^{-11}|e|a_B^2 \frac{Q_W}{N}$, leading to $Q^{\rm Cs}_W = -72.58(29)_{\rm exp.}(32)_{\rm th.}$. By comparing the experimental value with the up-to-date SM prediction in Eq.~(\ref{eq:QWSM}), a difference of $1.5\, \sigma$ is found,
$\delta Q^{\rm Cs}_{W} \equiv Q^{\rm Cs}_W-Q_{W}^{\textrm {SM+rad. corr.}} = 0.65(43)$.  
This translates in a similar deviation in the weak mixing angle, giving 
$\sin^2\theta_W = 0.2356(20)$, to be compared to the SM value in Eq.~(\ref{sin2theta}).\\


In this paper, we want to discuss the effect on $Q^{\rm Cs}_{W}$ of the
difference between the neutron and proton distribution in a nucleus, in view of a recent measurement of the average neutron rms distribution radius, $R_n$, of $^{133}\text{Cs}$ and $^{127}\text{I}$~\cite{PhysRevLett.120.072501}. Indeed, the parity violation in atoms is dominated by the $Z$-boson exchange between atomic electrons and neutrons, and so $({\rm Im}\, E_{\rm PNC})_{\rm th.}$ must be computed from the atomic wave functions. Since the wave function of the atomic electrons varies over the dimension of the nucleus, the final electroweak interaction with the nucleons depends on the spatial distribution of both protons and neutrons~\cite{PhysRevC.46.2587}. The effect of the different neutron and proton distributions has been explicitly considered in the atomic theory calculations in Ref.~\cite{PhysRevD.45.1602}, but at the end it was neglected, and the same distribution for protons and neutrons was assumed, because the estimated size of the correction was small compared to existing uncertainties at the time. However, as the experimental accuracy improved, it was realised that the effect could have no longer be neglected. 
Following the notation introduced in Ref.~\cite{Pollock:1999ec}, the effect of the finite nuclear size is to modify $N$ and $Z$ in Eq.~(\ref{QWSMEWcorrections1}) to $q_{{n}} N$ and $q_{{p}} Z$ respectively, where
\begin{equation}
q_{{n} ({p})} = \int f(r)
\rho_{{n} ({p})}(r) d^{3} r \,.
\label{weakcharge}
\end{equation}
Here $f(r)$ is a q-independent folding function determined from the radial dependence of the electron axial transition matrix element inside the nucleus, while $\rho_{{n} ({p})}$ is the neutron (proton) spatial distribution normalized to unity. The difference
between $q_n$ and $q_p$ has the effect of
modifying the effective weak charge SM value as\footnote{There are additional small corrections to $\Delta Q_W^{n-p}$ arising from the internal structure of the nucleon, but these can be safely neglected~\cite{PhysRevC.46.2587}.}
\begin{equation}
Q_W^{\rm SM+rad. corr.+n.s.}=Q_W^{\rm SM+rad. corr.} + \Delta
Q_W^{n-p}\,,
\label{weakchargeii}
\end{equation}
where the correction, $\Delta Q_W^{n-p}$, is defined as
\begin{equation}
\Delta Q_W^{n-p} = N(1-q_n/q_p)\,.
\label{chargeshift}
\end{equation}

An estimate of the
effect of different possible neutron distributions on $\Delta
Q_W^{n-p}$ can be obtained assuming a uniform
nuclear charge distribution ($\rho(r)$ constant out to some
radius), and then parametrizing the neutron distribution only with the value of $R_{{n}}$~\cite{PhysRevLett.65.2857}.
Using this approximation, one solves the Dirac equation
for the electron axial matrix elements, $f(r)$, near the origin by
expanding in powers of $\alpha$.
Finally, assuming
$R_n \approx R_p$, being $R_p$ the well-experimentally known proton rms distribution radius, and introducing a single small parameter, $\epsilon\equiv (R_n^2/ R_p^2)-1$, it is possible to find~\cite{PhysRevC.46.2587,Pollock:1999ec,Horowitz:1999fk}
\begin{eqnarray}
q_p \approx & 1-(Z\alpha)^2(0.26),\\
\label{approxi}
q_n(R_{{n}}) \approx \,& 1-(Z\alpha)^2(0.26+0.221\epsilon),\\
\label{approxii}
\Delta Q_W^{n-p}(R_n) \approx & N (Z\alpha)^2 (0.221
\epsilon)/q_p .
\label{approxiii}
\end{eqnarray} 

In Eq.~(\ref{approxii}) it is shown the rough dependence of the
weak charge correction on the difference between neutron and proton
distributions, characterized by $\epsilon$.

In the most recent re-evaluation of the APV result already cited~\cite{PhysRevLett.109.203003}, the authors, among other corrections, attempted to remove the effect of the difference between $R_n$ and $R_p$, originally calculated in Ref.~\cite{PhysRevA.65.012106}, from the experimental value of $Q_W^{\rm Cs}$. This allowed a direct comparison with the SM prediction $Q_W^{\rm SM+rad. corr.}$, that does not include this effect. Since at the time of Ref.~\cite{PhysRevLett.109.203003} there wasn't any caesium neutron radius measurement, the correction on $({\rm Im}\, E_{\rm PNC})_{\rm th.}$ due to the difference between $R_n$ and $R_p$ has been obtained exploiting antiprotonic atom x-ray data~\cite{PhysRevLett.87.082501}. From these data, the value of the so-called neutron skin, $\Delta R_{np}\equiv R_{n} - R_{p}$, has been
measured for a number of elements. From a fit to these measurements, the extrapolated neutron skin value for each element was found to be
$\Delta R_{np}=(-0.04\pm 0.03)+\left(1.01\pm 0.15\right)\frac{N-Z}{A}\textrm{ fm}$, where A is the mass number. For $^{133}\text{Cs}$, one can extrapolate the value $\Delta R_{np}=0.13(4)\textrm{ fm}$, which, combined with the very well known value of $R_{p}=4.807(1) \textrm{ fm}$ for $^{133}\text{Cs}$ at the time~\cite{JOHNSON1985405}, gave a correction to $({\rm Im}\, E_{\rm PNC})_{\rm th.}$ of \mbox{$-0.0018(5)\times 10^{-11}|e|a_B^2 \frac{Q_W}{N}$}, as explicitly visible in table IV of Ref.~\cite{PhysRevLett.109.203003}.  
Here, we want to remove this correction (but keeping all the other corrections introduced in Ref.~\cite{PhysRevLett.109.203003}), in order to retrieve the experimental value of $Q_W^{\rm Cs}$ including the neutron skin effect, indicated with $Q_W^{\rm Cs \,n.s.}$. Removing this correction, the theoretical amplitude needed in Eq.~(\ref{QWeq}), becomes
\begin{equation}
\label{MyImEPNC}
({\rm Im}\, E_{\rm PNC})_{\rm th.}^{\rm n.s.}=0.8995(40)\times10^{-11}|e|a_B^2 \frac{Q_W}{N}\,,
\end{equation}
from which one finds $Q_W^{\rm Cs\,n.s.}=-72.44(43)$.

In Ref.~\cite{PhysRevLett.120.072501}, using the coherent elastic neutrino-nucleus scattering (CEnNS) data of the COHERENT experiment~\cite{Akimov:2017ade}, the authors were able to determine for the first time the average neutron rms radius of $^{133}\text{Cs}$ and $^{127}\text{I}$. Being the values of $R_{p}$ for these nuclei very similar, with a difference of about 0.05 fm~\cite{Fricke:1995zz}, it is expected that also the values of $R_{n}$ are equal within the current uncertainties. They obtained a practically model-independent value of $R^{\rm COHER}_{n}=5.5{}^{+0.9}_{-1.1}\,\text{fm}$ that corresponds to a neutron skin value of $\Delta R^{\rm COHER}_{np} \simeq 0.7{}^{+0.9}_{-1.1}\,\text{fm}$. 

This provides a unique opportunity to derive for the first time a data-driven correction to $Q_W^{\rm Cs\,n.s.}$, in order to obtain a new value of $Q_W^{\rm Cs}$ that can be directly compared with $Q_W^{\rm SM+rad. corr.}$. Using Eq.~(\ref{approxiii}) one finds
\begin{eqnarray}\nonumber
Q_W^{\rm Cs}&=&
Q_W^{\rm Cs\,n.s.} - \Delta
Q_W^{{n-p}} \\ 
&=& -72.44(43)-0.9{}^{+1.2}_{-1.5}=-73.3{}^{+1.3}_{-1.6}\,,
\end{eqnarray}
which now relies on a direct experimental input for a caesium nucleus. 
We note that now the central value of $Q^{\rm Cs}_W$ is in better agreement with the SM value in Eq.~(\ref{eq:QWSM}). Indeed, the difference with the SM prediction is now $\delta Q^{\rm Cs}_{W}=-0.1{}^{+1.3}_{-1.6}$. The effect of the inclusion of the experimental input of the neutron distribution radius of the caesium has the effect to shift the central value, but also to increase significantly its uncertainty. 
Relating this result to the weak mixing angle~\cite{PhysRevLett.109.203003}, we obtain\footnote{In the determination of $R_n$ for caesium using CEnNS data the value of the weak mixing angle is used, however this dependence is very small and it does not change the outcome of this paper.} the new APV value of $\sin^2 \vartheta_{\text{W}}=0.239{}^{+0.006}_{-0.007}$, with a central value in very good agreement with the SM at low momentum transfer, as showed by the red point in Fig.~\ref{fig:running}.\\

Since the APV measurement depends so crucially on the neutron skin value, the first can be used in combination with the COHERENT data to determine the value of $R_n$ for $^{133}\text{Cs}$.
Assuming that the SM is correct, and so assuming the PDG value~\cite{PDG2018} of the weak mixing angle at low momentum transfer\footnote{Note that this procedure is perfectly consistent since the neutron radius from COHERENT data has been obtained assuming the same low-energy value of the weak mixing angle.}, the following combined APV and COHERENT least-squares function can be built
\begin{align}
\chi^2
=
\null & \null
\sum_{i=4}^{15}
\left(
\dfrac{
N_{i}^{\text{exp}}
-
\left(1+\alpha\right) N_{i}^{\text{th}}
-
\left(1+\beta\right) B_{i}
}{ \sigma_{i} }
\right)^2
\nonumber
\\
\null & \null
+
\left( \dfrac{\alpha}{\sigma_{\alpha}} \right)^2
+
\left( \dfrac{\beta}{\sigma_{\beta}} \right)^2
+
\chi_{APV}^2
\,,
\label{chitot}
\end{align} 
where the first three terms refer to the COHERENT data analysis (see Ref.~\cite{PhysRevLett.120.072501} for details), in which in every energy bin $i$, $N_{i}^{\text{exp}}$
and $N_{i}^{\text{th}}$ are, respectively, the experimental and theoretical number of CEnNS events, with the latter depending on $R_n$, $B_{i}$ is the estimated number of background events from Ref.~\cite{Akimov:2017ade}, and $\sigma_{i}$ is the statistical uncertainty. The nuisance parameters
$\alpha$ and $\beta$ represent the systematic uncertainty of the signal rate and
of the background rate, respectively, with associated
standard deviations of $\sigma_{\alpha} = 0.28$ and $\sigma_{\beta} = 0.25$
\cite{Akimov:2017ade}.
The last ingredient in the least-squares function refers to the APV measurement for caesium and can be written as 
\begin{equation}
\chi_{APV}^2=
\left(
\dfrac{
(Q_W^{\rm Cs\,n.s.}- \Delta
Q_W^{{n-p}})
-
Q_W^{\rm SM+rad. corr.}
}{ \sigma_{APV} }
\right)^2\,,
\label{chiAPV}
\end{equation}
in which $\sigma_{APV}$ is the total uncertainty corresponding to $\sigma_{APV}=0.43$. 
 
\begin{figure}[!t]
\centering
\includegraphics*[width=\linewidth]{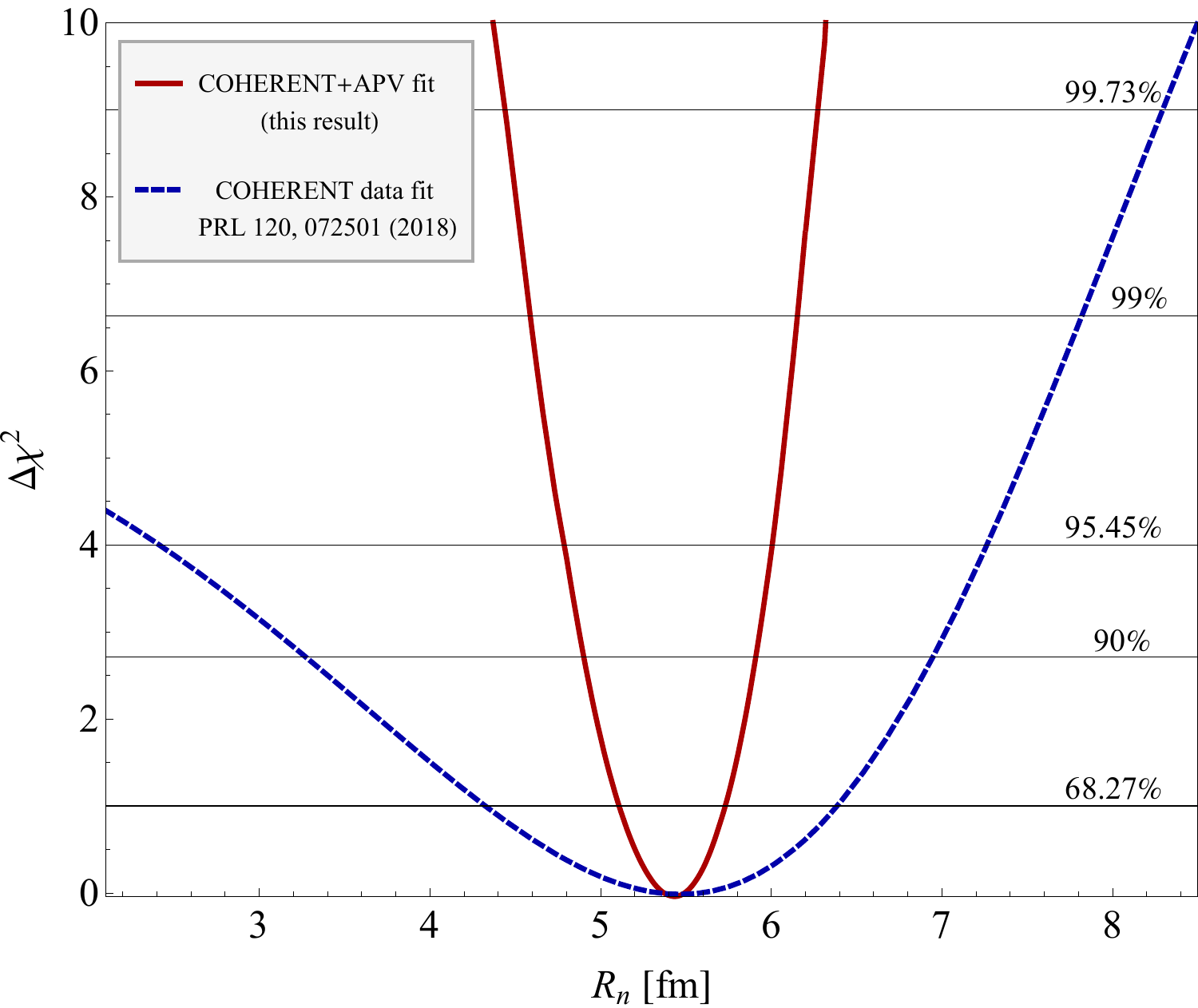}
\caption{ \label{fig:deltachi2} With the red solid curve it is shown the $\Delta \chi^2=\chi^2-\chi_{\rm min}^2$, with $\chi^2$ as defined in Eq.~(\ref{chitot}), as a function of the neutron rms radius, $R_{n}$, obtained from the combined fit of the COHERENT data and the APV caesium measurement. With the blue dashed line it is shown the $\Delta \chi^2$ obtained fitting the COHERENT data alone (see Ref.~\cite{PhysRevLett.120.072501}).
} 
\end{figure}
Figure~\ref{fig:deltachi2} shows the corresponding marginal values of the $\chi^2$
as a function of $R_{n}$, superimposed with the values obtained fitting the COHERENT data alone. 
One can see that the inclusion of the APV measurement allows to shrink significantly the $\Delta\chi^2$ profile, and to make it symmetric with respect to the best fit value, reducing the available space for low $R_n$'s. The result with the inclusion of the APV measurement is
\begin{equation}
R_{n}=5.42\pm 0.31\,\text{fm}\,,
\label{rn}
\end{equation}
which is highly compatible with that obtained using CEnNS data only. 
Using the value found in Eq.~(\ref{rn}) and the updated value of $R_{p}$ for $^{133}\text{Cs}$, $R_{p}=4.804(1) \,\text{fm}$~\cite{Fricke:1995zz}, it is possible to infer for the first time a meaningful value of the $^{133}\text{Cs}$ neutron skin, which is 
\begin{equation}
\Delta R_{np}=0.62\pm 0.31\,\text{fm}\,.
\label{nskin}
\end{equation}
The central value shows a preference for a possible larger than the model-predicted values~\cite{Horowitz:1999fk,PhysRevLett.120.072501}, despite more precise measurements are required to confirm it.

Information on $\Delta R_{np}$ is of particular importance because it is correlated with many neutron-rich matter properties, like the the total and the isovector densities~\cite{Brown:2000pd,Horowitz:2000xj,Reinhard:2010wz,Tsang:2012se,Hagen:2015yea}. 
Moreover, the value of $\Delta R_{np}$ provides important information on the Equation Of State (EOS) of nuclear matter. Recently, this field of research has gained great attention after the observation of a binary neutron star inspiral performed by the Advanced LIGO and the Advanced Virgo gravitational-wave experiments~\cite{TheLIGOScientific:2017qsa}, from which it was possible to infer information on neutron star EOS~\cite{Abbott:2018exr, Fattoyev:2017jql}.
In particular, one of the main features of EOS, that has been recently studied in detail, is the symmetry energy, $S$, and its density dependence, a quantity known as the slope parameter $L=3\rho_{0}\left(\dfrac{dS}{d\rho}\right)_{\rho_0}$, being $\rho_{0}$ the so-called saturation density.
Since the neutron skin is linearly correlated with $L$, an experimental measure of $\Delta R_{np}$ is probably the most reliable way to determine $L$ (see Ref.~\cite{Baldo:2016jhp} for a review). Moreover, the value of $L$ has a direct impact on the properties of neutron stars. Indeed, larger values of $L$, and thus larger values of $\Delta R_{np}$, would suggest a stiffer EOS and allow for even larger neutron star masses. 
Finally, a more precise determination of $R_{n}$ is crucial in order to better constraint~\cite{,Cadeddu:2018sfb} the most insidious background for future direct dark matter detectors~\cite{Billard:2013qya} which is due to CEnNS process. Indeed, if not well characterized, it will crucially limit the dark matter discovery potential. 
Future dark matter experiments like XENONnT~\cite{Aprile:2015uzo}, DARWIN~\cite{Aalbers:2016jon} and LZ~\cite{Akerib:2018lyp} will use xenon as a target material and since caesium has similar atomic and mass numbers to that of xenon, information on $R_n$ found in this paper could help to better estimate the total number of expected CEnNS background events.\\

In conclusion, using the model independent average neutron rms radius of $^{133}\text{Cs}$ and $^{127}\text{I}$ obtained in Ref.~\cite{PhysRevLett.120.072501} from the analysis of CEnNS data of the COHERENT experiment, we show that the long-standing $1.5\,\sigma$ deviation of the weak mixing angle measurement at very low momentum transfer, obtained via APV effects in caesium, is weaken. The new value of the EW nuclear charge $Q^{\rm Cs}_W =-73.3{}^{+1.3}_{-1.6}$ is in very good agreement with the SM value, with a difference $\delta Q^{\rm Cs}_{W}=-0.1{}^{+1.3}_{-1.6}$. This corresponds to a new value of the weak mixing angle of $\sin^2 \vartheta_{\text{W}}=0.239{}^{+0.006}_{-0.007}$.  However, the uncertainty, which is dominated by that of the caesium neutron distribution radius, is significantly enlarged with respect to the previous one, that on the contrary did not use any direct experimental input. This result highlights the crucial importance of pursuing more precise measurements of $R_n$, for instance using future updates of COHERENT and other CEnNS experiments, as showed in Ref.~\cite{PhysRevLett.120.072501}.  

Finally, exploiting the fact that the APV result largely depends on the caesium neutron radius and assuming that the SM is correct, we perform a combined fit of the latter with the COHERENT data. A better determination of the neutron rms radius of caesium is obtained, namely $R_{n}=5.42\pm 0.31\,\text{fm}$. The inclusion of the APV data allows to reduce the uncertainty on the radius by one third with respect to the first experimental result in Ref.~\cite{PhysRevLett.120.072501} and a meaningful determination of the neutron skin of caesium nuclei, leading to $\Delta R_{np}=0.62\pm 0.31\,\text{fm}$, showing for the first time a $2\, \sigma$ deviation from zero.\\


M. Cadeddu is grateful to C. Giunti and M. Lissia for stimulating discussions.

\nocite{*}
\bibliographystyle{apsrev4-1}
\bibliography{bibAPV}

\end{document}